\documentclass[smallcondensed,final,a4paper]{svjour3}
\smartqed  
\usepackage{graphicx}
\usepackage{graphics}
\usepackage{epstopdf}
\usepackage{subfigure}
\usepackage{longtable}
\usepackage{pdfpages}
\usepackage{pgfplotstable}
\usepackage{changepage}
\usepackage{textcomp}
\usepackage{wrapfig}
\usepackage[numbers,compress]{natbib}
\usepackage{mathptmx}      
\journalname{Theoretical and Computational Fluid Dynamics}
\begin{document}

\title{Waves in strong centrifugal fields: dissipationless gas
}


\author{S.V. Bogovalov, V.A. Kislov,  I.V. Tronin 
}


\institute{S. Bogovalov \at
              National Research Nuclear University ``MEPhI'', Kashirskoje shosse, 31, Moscow, 115409, Russia \\
              Tel.: +7-499-323-9045\\
              Fax: +7-495-324-21-11\\
              \email{svbogovalov@mephi.ru}           
}

\date{Received: date / Accepted: date}

\maketitle

\begin{abstract}
Linear waves  are investigated  in a rotating gas under the condition of strong centrifugal acceleration of the order $10^6 g$ realized in gas centrifuges for separation of uranium isotopes. Sound waves splits into three families of the waves  under these conditions. Dispersion equations are obtained. The characteristics of the waves strongly differ from the conventional sound waves on polarisation, velocity of propagation and distribution of energy of the waves in space for two families 
having frequencies above and below the frequency of the conventional sound waves.  The energy of these waves  is localized in rarefied region of the gas. The waves of the third family were not specified before. They propagate exactly along the rotational axis with the conventional sound velocity.  These waves are polarized only along the rotational axis. Radial and azimuthal motions are not excited. Energy  of the  waves is concentrated near the wall of the rotor where the density of the gas is largest. 
\end{abstract}

\section{Introduction}

Separation of heavy isotopes in strong centrifugal fields is used for industrial production of enriched uranium.  Rotors of gas centrifuges (GC) rotate with 
linear velocity a few hundreds meters per second \citep{iguasu}.  The centrifugal acceleration can reach $\sim 10^6 g$ at the radius of the rotor of a few  centimeters . Nevertheless, strong centrifugal field is not the only factor which results into efficient isotope separation in GC. A range of physical processes are explored to produce  an additional axial circulation which essentially increases the efficiency of separation of the GC. This secondary circulation is one of the key feature of the industrial GC. 

\begin{figure}{r}
\begin{center}
 \includegraphics[width=0.92\linewidth]{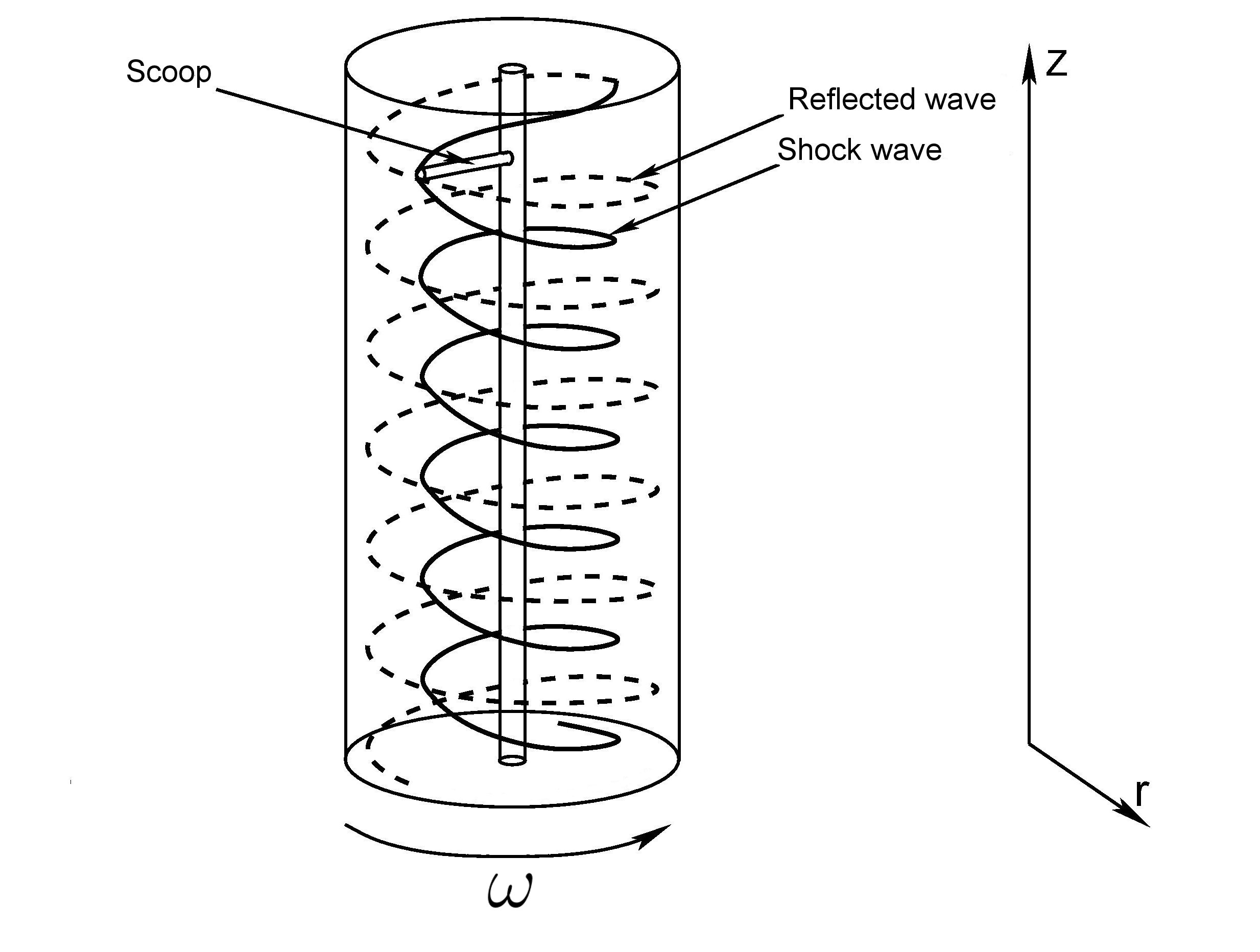}
\end{center}
\caption{Scheme of the gas centrifuge with scoops. Solid line - shock wave produced by the scoop, dashed line - the wave reflected from the upper end cap.}
\label{fig:model}
\end{figure}

A pair of scoops located near the end caps of the GC are used to remove enriched and depleted gas mixture from the GC. Simultaneously, they provide  the additional axial circulation  due to the mechanical brake of the gas. Mach number of the gas in GC is of the order ~7 because the sound velocity of the working gas $UF^6$ is of the order $86~ \rm m/s$ at room temperatures. Interaction of the gas with the scoop is accompanied by formation of strong shock wave which propagates along the rotational axis. Fig. \ref{fig:model} schematically shows that the shock wave forms a spiral wave propagating from one end of the GC to another. It can reflect from the end caps  of the GC  forming  waves running in both directions along the rotational axis. The amplitude of the shock waves is damped rather quickly. In the largest part of the rotor they propagate as small amplitude waves.  Therefore it is reasonable to consider them firstly in a linear approximation.

It is well known that the waves can produce the flow of gas due to their absorption \citep{br}. These are so-called acoustic flows described by Lord Relei \citep{rel1,rel2}.  They are produced  due to transfer of the energy and momentum from the waves to the gas due to the molecular viscosity.  
This way the waves generated by the scoops can provide an additional mechanism of generation of the secondary axial circulation in GC. This mechanism can essentially differ  from the conventional mechanism of the circulation generation. The waves can propagate and therefore transfer the breaking torque at larger 
distance from the scoops than it happens at the axisymmetric  brake.  Exploration of this mechanism can change  efficiency of the isotope separation and a working parameters of the GC.

The axial secondary circulation in the GC is defined by the spatial distribution of the rate of absorption of the waves. Application of the conventional equations to the calculation of the  rate for dumping of the waves is not possible on several reasons in this case. Firstly, strong centrifugal acceleration achieving $10^6 g$ dramatically changes the characteristics of the linear waves. Secondly, the strong radial density gradient changing on 6 orders of magnitude per  radius variation in $\sim 1 cm$, gives the rate of absorption changing also on 6 orders of magnitude. On these reasons it is impossible even to estimate the length of the wave propagation.     

This paper is devoted to the properties of the waves in strong centrifugal fields. This is not new problem. Propagation of the linear waves in the rotating gas have been investigated in many works before. Firstly, it is well known that there is an additional type of waves named inertial  produced in the rotating fluid by the Coriolis forces \citep{nf1,nf2,nf3}. These waves has been widely considered in application to physics of oceans and atmosphere  \citep{oa1,oa2,oa3,oa4}.  The stability of the rotating fluids has been investigated in  \citep{sg1,sg2,sg3,sgn2,sgn3}. Nevertheless, all these studies concerned incompressible fluid or gases in relatively weak centrifugal fields when the gradient of density along radial direction can be neglected \citep{sgn1,rotation1,rotation2}.  We consider the case typical for the GC used for separation of isotopes when the density changes on many orders of magnitude along the radius.  

Strong centrifugal field of the GC makes the analysis of the flow of the gas complicated problem.  Density of the gas in typical GC varies  12-14 orders of magnitudes along radius from the wall of the rotor to the axis of rotation.  In this conditions application of Navier - Stocks equations to the problem is possible only in the narrow region 
with size $1-2 cm$ from the rotor  wall (typical rotor radius is $6-8 cm$). At smaller radius the gas is getting so rare that the path length of the molecules  of the gas exceeds the rotor radius. 
Hydrodynamic equations are not valid here. Kinetic methods should be used in the so called Knudsen region to describe the gas flow \citep{borman}.

The main objective of this paper is to understand what kinds of waves are possible in the gas subjected to the centrifugal acceleration typical for GC used for separation of the uranium isotopes. Therefore, we use here a simplified model of the gas assuming that the hydrodynamic equations  are valid everywhere. Moreover,   
the case of dissipationless gas is under the consideration. This means that the heat conductivity and molecular viscosity are neglected. In next papers we consider the role of the  dissipative processes and corrections into the model introduced by the presence of the rarefied gas zones where the hydrodynamic approximation is violated.

\section{Wave equation in strong centrifugal field}

Let us consider an ideal gas having molar mass
 $M$  and rotating with the angular velocity  $\omega$. The system of equations defining the dynamics of the gas in the rotating frame system is as follows  \citep{landau}:
\begin{equation}
 \frac{\partial D}{\partial t}+\frac{\partial \left( r D v_r \right)}{r \partial r}+\frac{\partial \left( D v_\phi \right) }{r \partial \phi}+\frac{\partial \left( D v_z \right) }{\partial z}=0,
\label{eq1}
\end{equation}
\begin{equation}
 D \frac{\partial v_r}{\partial t}+D \left( v_r \frac{\partial v_r}{\partial r}+\omega \frac{\partial v_r}{\partial \phi}+v_\phi\frac{\partial v_r}{r \partial \phi}+v_z\frac{\partial v_r}{\partial z}-\omega^2 r-2 \omega v_\phi -\frac{v^2_\phi}{r}\right)=-\frac{\partial P}{\partial r},
\end{equation}
\begin{equation}
 D \frac{\partial v_\phi}{\partial t}+D \left( v_r \frac{\partial v_\phi}{\partial r}+\omega \frac{\partial v_\phi}{\partial \phi}+v_\phi \frac{\partial v_\phi}{r \partial \phi}+v_z\frac{\partial v_\phi}{\partial z}+2\omega v_r +\frac{v_\phi v_r}{r}\right)=-\frac{\partial P}{r \partial \phi},
\end{equation}
\begin{equation}
 D \frac{\partial v_z}{\partial t}+D \left( v_r \frac{\partial v_z}{\partial r}+\omega \frac{\partial v_z}{\partial \phi}+v_\phi \frac{v_z}{r \partial \phi}+v_z\frac{\partial v_z}{\partial z}\right)=-\frac{\partial P}{\partial z},
\end{equation}
\begin{eqnarray}
 D c_p \frac{\partial \tau}{\partial t}+D c_p \left( v_r \frac{\partial \tau}{\partial r}+\omega \frac{\partial \tau}{\partial \phi}+v_\phi\frac{\partial \tau}{r \partial \phi}+v_z\frac{\partial \tau}{\partial z}\right)=& \nonumber\\
=\frac{\partial P}{\partial t}+v_r \frac{\partial P}{\partial r}+\omega \frac{\partial P}{\partial \phi}+v_\phi \frac{\partial P}{r \partial \phi}+v_z \frac{\partial P}{\partial z}.
\label{eq5}
\end{eqnarray}
where $c_p$ -- is  the specific heat capacity for constant pressure, $P$- pressure, $D$ - density, $\tau$ - temperature  and $v_r,v_\phi, v_z$  are the radial, azimuthal and axial components of the velocity.

 Parameters of the gas can be presented as the sum of parameters of the rigid body rotation and some deviation
\begin{equation}
 v_r=\bar{v_r},v_z=\bar{v_z}, v_\phi=\bar{v_\phi},P=p_0+\bar{p}, D=\rho_0+\bar{\rho}, \tau=T_0+\bar{T},
\end{equation}
where $p_0, \rho_0, T_0$ -- are the rigid body rotation pressure, density and temperature, respectively. $\bar{v_r}, \bar{v_\phi}, \bar{v_z}, \bar{p}, \bar{\rho}, \bar{T}$ -- are deviations of radial, azimuthal, axial velocity, pressure, density and temperature from the rigid body rotation values, respectively.

Dependence of the rigid body pressure and density on the radius has a form 
\begin{equation}
p_0\left( r\right)=p_w \exp(\frac{M \omega^2}{2RT_0}\left( r^2-a^2\right)) ,
\end{equation}
\begin{equation}
 \rho_0\left( r\right)=\rho_w \exp(\frac{M \omega^2}{2RT_0}\left( r^2-a^2\right)),
\end{equation}
where $a$ -- is the radius of the rotor, $p_w$ and $\rho_w$ -- are the pressure and density of the gas at the wall of the rotor, respectively.

For the ideal gas 
\begin{equation}
 \rho=\frac{M p}{RT}.
\end{equation}
For the deviation from rigid body rotation we have 
\begin{equation}
 \bar{\rho}=\frac{M\bar{p}}{RT_0}-\frac{\rho_0 \bar{T}}{T_0}.
\label{form:rhoot}
\end{equation}

The waves produced by the scoopes are not axisymmetric. They form a spiral. Nevertheless, 
it is reasonable to consider the simplest case of the axisymmetric waves as the first step in solution of the general problem.
The terms with derivative  $\frac{\partial}{\partial \phi}$ can be neglected in this case.  Linearization of eqs. (\ref{eq1} - \ref{eq5}) gives
\begin{equation}
 \frac{\partial \bar{\rho}}{\partial t}+\frac{\partial \left(r \rho_0 \bar{v_r}\right)}{r \partial r}+\frac{\partial \rho_0 \bar{v_z}}{\partial z}=0,
\end{equation}
\begin{equation}
 \rho_0 \frac{\partial \bar{v_r}}{\partial t}-2 \rho_0 \omega \bar{v_\phi}-\bar{\rho}\omega^2 r=-\frac{\partial \bar{p}}{\partial r},
\end{equation}
\begin{equation}
 \rho_0 \frac{\partial \bar{v_\phi}}{\partial t}+2\rho_0 \omega \bar{v_r}=0,
\end{equation}
\begin{equation}
 \rho_0\frac{\partial \bar{v_z}}{\partial t}=-\frac{\partial \bar{p}}{\partial z},
\end{equation}
\begin{equation}
 \rho_0 c_p \frac{\partial \bar{T}}{\partial t}=\frac{\partial \bar{p}}{\partial t}+\rho_0 \omega^2 r \bar{v_r}.
\end{equation}

Let us assume that periodical along axial direction $z$ wave has the wave vector $k$ and frequency  $\Omega$.  All the perturbations take a form
\begin{eqnarray}
 \bar{v_r}=V_r e^{-i \Omega t+i k z}, \bar{v_\phi}& = &V_\phi e^{-i \Omega t+i k z}, \bar{v_z}=V_z e^{-i \Omega t+i k z},\nonumber\\ 
\bar{p}=p e^{-i \Omega t+i k z}, \bar{\rho}& = &\rho e^{-i \Omega t+i k z},\bar{T}=T e^{-i \Omega t+i k z}.\nonumber
\end{eqnarray}
Then we obtain the following system of equations
\begin{equation}
 -i \Omega \rho+\frac{\partial \left(r \rho_0 V_r\right)}{r \partial r}+i k \rho_0 V_z=0,
\label{form:nepr}
\end{equation}
\begin{equation}
 -i \Omega \rho_0 V_r -2 \rho_0 \omega V_\phi-\rho \omega^2 r= -\frac{\partial p}{\partial r},
\label{form:radial}
\end{equation}
\begin{equation}
 -i \rho_0\Omega V_\phi+2 \rho_0 \omega V_r=0,
\label{form:azim}
\end{equation}
\begin{equation}
 \rho_0 \Omega V_z= k p,
\label{form:axial}
\end{equation}
\begin{equation}
 -i \rho_0 c_p \Omega T=-i \Omega p +\rho_0 \omega ^2 r V_r.
\label{form:en}
\end{equation}

Eqs. (\ref{form:azim}) and  (\ref{form:axial}) give
\begin{equation}
 V_\phi=\frac{2 \omega V_r}{i \Omega},
\label{form:vph}
\end{equation}
\begin{equation}
 V_z=\frac{k p}{\Omega \rho_0}.
\label{form:vz}
\end{equation}

Eq. (\ref{form:en}) gives the following expression for the perturbation of the temperature
\begin{equation}
 T=\frac{-i \Omega p+\rho_0 \omega^2 r V_r}{-i \rho_0 c_p \Omega}.
\label{form:ottemp}
\end{equation}

Substitution of eq. (\ref{form:ottemp}) into (\ref{form:rhoot}) gives
\begin{equation}
 \rho=\frac{p}{c^2}+\frac{\rho_0 \omega^2 r V_r}{i \Omega c_p T_0}.
\label{form:rho}
\end{equation}
Here the following equation is used
\begin{equation}
 \frac{M}{RT_0}-\frac{1}{c_p T_0}=\frac{c_p M -R}{c_p R T_0}=\frac{c_v M}{c_p RT_0}=\frac{1}{c^2},
\end{equation}
where $c_v$ -- is the specific heat capacity for constant volume.

\begin{figure}
\centering
\includegraphics[width=0.5\linewidth]{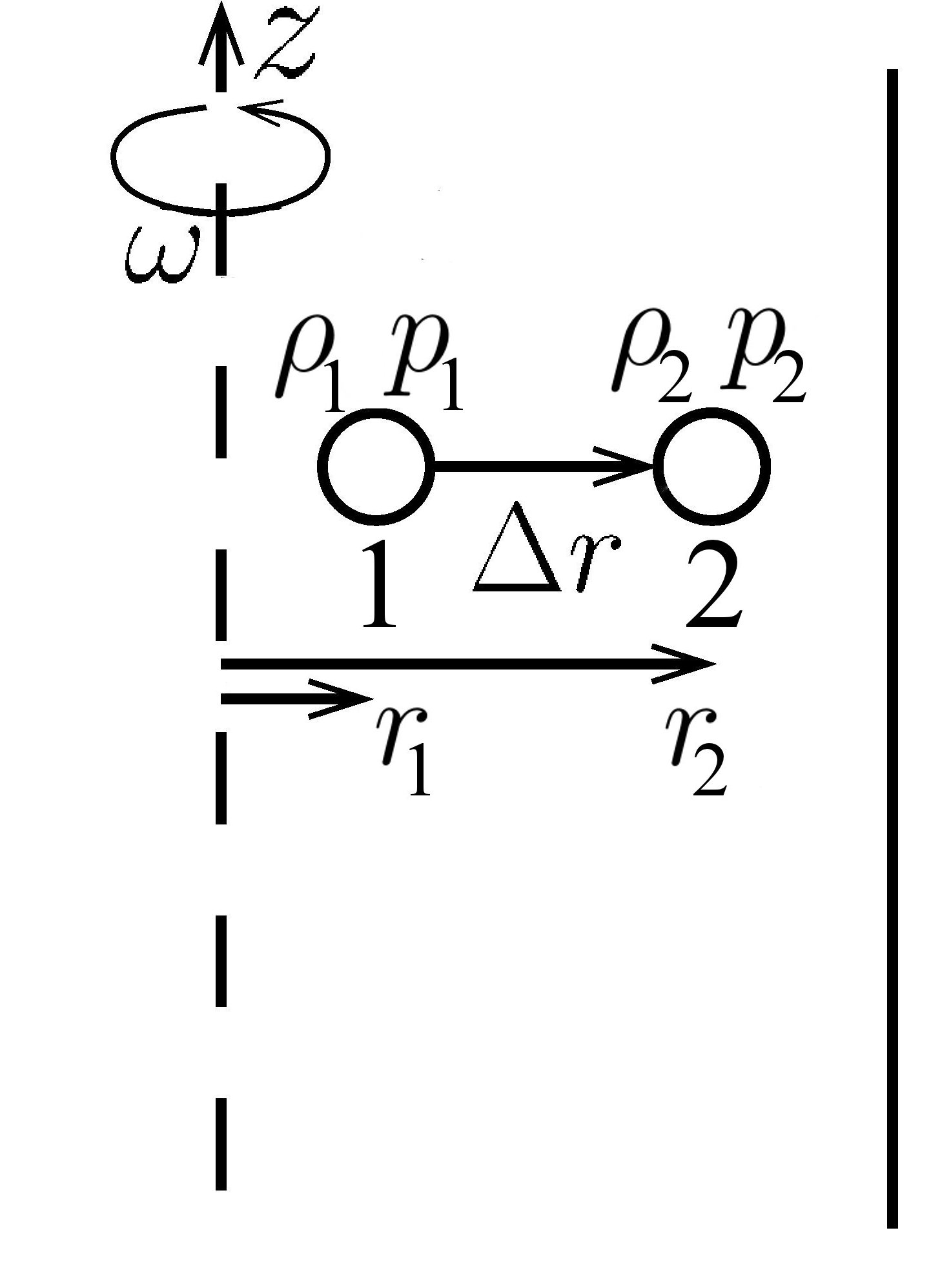}
\caption{Displacement of the piece of the gas from position 1 into position 2 is accompanied by the work of the centrifugal force which 
modifies the conventional adiabatic relationship between density and pressure.}
\label{fig:displ}
\end{figure}

It is worth to pay attention that eq. (\ref{form:rho}) differs from the conventional relationship for the adiabatic variation of the pressure and density ${\bar p}=c^2{\bar \rho}$. The origin of the 
last term in eq.  (\ref{form:rho}) is connected with the work of the centrifugal field. To understand the nature of this term let us consider the process of motion of a small 
piece of the gas from the initial position 1 in the initial moment 1  to the position 2 separated by the radial distance $\Delta r$ at the next time moment 2 as it is shown in fig. \ref{fig:displ}. The conventional adiabatic equation has a form
\begin{equation}
\rho_2= \rho_1({p_2\over p_1})^{{1\over \gamma}}, 
\end{equation}
where $\gamma=C_p/C_v$ , $p_1, \rho_1$ are the pressure and density at position 1, $p_2, \rho_2$ are the pressure and density at position 2, respectively. In the uniform gas $p_1$ and $\rho_1$ are constants and we obtain the conventional equation for variation of the pressure and density
\begin{equation}
 {\delta \rho} = \frac{\delta p}{c^2},
\end{equation}
where $\delta \rho$ and $\delta p$ are variations of the density and pressure at the position 2 during time interval $\Delta t$.

In the strong centrifugal field the variation of the unperturbed state of the gas  with $r$ should be taken into account. This gives
\begin{equation}
 p_2={\delta p}+p_1+\frac{\partial p_0}{\partial r}\Delta r, \rho_2={\delta \rho}+\rho_1+\frac{\partial \rho_0}{\partial r}\Delta r.
\end{equation}

In this case the  connection between variations of the density and pressure takes a form
\begin{equation}
 {\delta \rho}={{\delta p}\over c^2}-{\partial \rho_0\over \partial r}\Delta r+ {1\over \gamma} {\partial \rho_0\over \partial r}\Delta r.
\end{equation}
This gives
\begin{equation}
\delta \rho={\delta p\over c^2}-{(\gamma-1)\over \gamma}{\rho_0 M\omega^2 r\over RT_0} V_r\Delta t,
\end{equation}
which is equivalent to eq.  (\ref{form:rho}). In the works by \citep{rotation1} and \citep{rotation2} this effect has been fully neglected. This is one of the reasons why their results can not be applied to the waves in GC.

Let us insert eqs. (\ref{form:vph}) and (\ref{form:rho}) into (\ref{form:radial}). Then we have
\begin{equation}
 \left( -i \Omega \rho_0 - \frac{4 \rho_0 \omega^2}{i \Omega}-\frac{\rho_0 \omega^4 r^2}{i \Omega c_p T_0}\right)V_r=\frac{\omega^2 r p}{c^2}-\frac{\partial p}{\partial r}.
\label{form:eq1}
\end{equation}

Substitution of eqs. (\ref{form:vph}), (\ref{form:vz}) and (\ref{form:rho}) into (\ref{form:nepr}) gives
\begin{equation}
 \left( \frac{i k^2}{\Omega}-\frac{i \Omega}{c^2}\right)p-\frac{\rho_0 \omega^2 r}{c_p T_0}V_r+\frac{\partial (r \rho_0 V_r)}{r\partial r}=0.
\label{form:eq2}
\end{equation}

Let us introduce the following notation
\begin{equation}
 A=\frac{i k^2}{\Omega}-\frac{i \Omega}{c^2},
\label{eqa}
\end{equation}
and express from eq. (\ref{form:eq2}) the perturbation of pressure as follows
\begin{equation}
 p=\frac{1}{A}\left( \frac{\rho_0 \omega^2 r V_r}{c_p T_0}-\frac{\partial (r \rho_0 V_r)}{r \partial r}\right).
\label{form:pres}
\end{equation}

Here we have to divide the right hand part of eq. (\ref{form:eq2}) on the factor $A$. This operation can be illegal under the condition $A=0$.
Therefore, this case we consider separately later.  

Substitution of eq. (\ref{form:pres}) into eq.  (\ref{form:eq1}) gives
\begin{eqnarray}
 \rho_0 \left( -i\Omega-\frac{4 \omega^2}{i \Omega}-\frac{ \omega^4 r^2}{i \Omega c_p T_0} \right) V_r =\frac{\omega^2 r}{A c^2}\left( \frac{\rho_0 \omega^2 r V_r}{c_p T_0}-\frac{\partial (r \rho_0 V_r)}{r\partial r}\right)-&\nonumber\\
-\frac{\partial}{A\partial r}\left(\frac{\rho_0 \omega^2 r V_r}{c_p T_0}-\frac{\partial(r \rho_0 V_r)}{r \partial r} \right).
\label{form:eq3}
\end{eqnarray}

Let us introduce
\begin{equation}
 B=-i \Omega -\frac{4 \omega^2}{i \Omega}.
\end{equation}
Then eq.  (\ref{form:eq3}) takes the form
\begin{eqnarray}
 \left( \rho_0 B -\frac{\rho_0 \omega^4 r^2}{i \Omega c_p T_0}\right)V_r=\frac{\omega^4 r^2 \rho_0 }{Ac^2 c_p T_0}V_r -\frac{\omega^2 \partial (r \rho_0 V_r)}{Ac^2 \partial r}-&\nonumber \\
\frac{\partial}{A\partial r}\left( \frac{\rho_0 \omega^2 r V_r}{c_p T_0}\right)+
\frac{\partial}{A\partial r}\left( \frac{\partial(r \rho_0 V_r)}{r \partial r}\right).
\end{eqnarray}
Introduction of new variable  $\Phi=\rho_0 V_r$ and multiplication of this equation on $A$ gives
\begin{equation}
A\left( B-\frac{ \omega^4 r^2}{i \Omega c_p T_0} \right)\Phi= \frac{\omega^4 r^2 }{c^2 c_p T_0}\Phi-\frac{\omega^2 M}{RT_0}\frac{\partial r \Phi}{\partial r}+\frac{\partial}{\partial r}\left( \frac{\partial(r \Phi)}{r \partial r}\right).
\end{equation}
Let us multiply now the equation on $r$, introduce new variable $Z=r\Phi$ and use new argument $x=r^2$. After that we obtain
\begin{equation}
 A\left( B-\frac{ \omega^4 r^2}{i \Omega c_p T_0} \right)Z=\frac{\omega^4 x}{c^2 c_p T_0}Z-\frac{2 \omega^2 M x}{RT_0}\frac{\partial Z}{ \partial x}+4x\frac{\partial^2 Z}{ \partial x^2}.
\end{equation}

It is convenient to introduce  $Z=Y e^{\frac{\omega^2 M x}{4RT_0}}$. After substitution into the equation above  it takes a form
\begin{equation}
  Y''+{1\over 4}\left(A'+{B\over x}'\right)Y=0,
\label{form:eqosn}
\end{equation}
where
\begin{equation}
 A'=-\frac{M^2 \omega^4}{4R^2 T^2_0}+\frac{k^2 \omega^4}{c_p T_0 \Omega^2},
\end{equation}
\begin{equation}
 B'=-\frac{\left(k^2-\frac{\Omega^2}{c^2} \right)\left( \Omega^2-4\omega^2\right)}{\Omega^2}.
\label{form:b}
\end{equation}
It is convenient to consider 
eq. (\ref{form:eqosn}) as the Shr\"{o}dinger equation for a particle in the potential of the form 
\begin{equation}
 U(x)=-{1\over 4}(A'+{B'\over x}) ~at ~x < a, ~U(x)=\infty, ~at  ~x \ge a.
\label{shredinger}
\end{equation}
with the energy eigenvalue equal to zero.
Boundary conditions for the equations follow from the boundary conditions for radial velocity $v_r$. It goes to zero 
at the axis of the rotor and at the wall of the rotor.  Therefore boundary conditions for $Y$ are 
\begin{equation}
 Y(0)=0,~~ Y(a^2)=0.
\end{equation}
This form of the equation with the boundary conditions allows us qualitatively understand  the behaviour of the solution.

\section{Solution}
\subsection{The case $A \ne 0$}

The solution of eq. (\ref{form:eqosn}) can be expressed as  \citep{abrstigun}
\begin{equation}
 Y=WM\left(\frac{B'}{4 \sqrt{A'}},0.5,\sqrt{A'} x\right),
\end{equation}
where $WM$ -- Whittaker function satisfying the condition $WM(x=0)=0$.
The boundary condition at $x=a^2$ provides us the 
dispersion relations for $\Omega(k)$.

Then all the perturbations are expressed through $Y$ as follows
\begin{equation}
 V_r=\frac{Y e^{\frac{\omega^2 M r^2}{4RT_0}}}{r \rho_0},
\label{eqvr}
\end{equation}
\begin{equation}
 V_\phi=\frac{2 \omega Y e^{\frac{\omega^2 M r^2}{4RT_0}}}{i \Omega r \rho_0},
\end{equation}
\begin{equation}
 V_z=\frac{k p}{\Omega \rho_0},
\label{eqvz}
\end{equation}
\begin{equation}
 p=\frac{1}{A}\left( \frac{ \omega^2  Y e^{\frac{\omega^2 M r^2}{4RT_0}}}{c_p T_0}-\frac{\partial (Y e^{\frac{\omega^2 M r^2}{4RT_0}})}{r \partial r}\right),
\end{equation}
\begin{equation}
 T=\frac{-i \Omega p+\rho_0 \omega^2 r V_r}{-i \rho_0 c_p \Omega}.
\end{equation}

\begin{figure}
\centering
\includegraphics[width=0.99\linewidth]{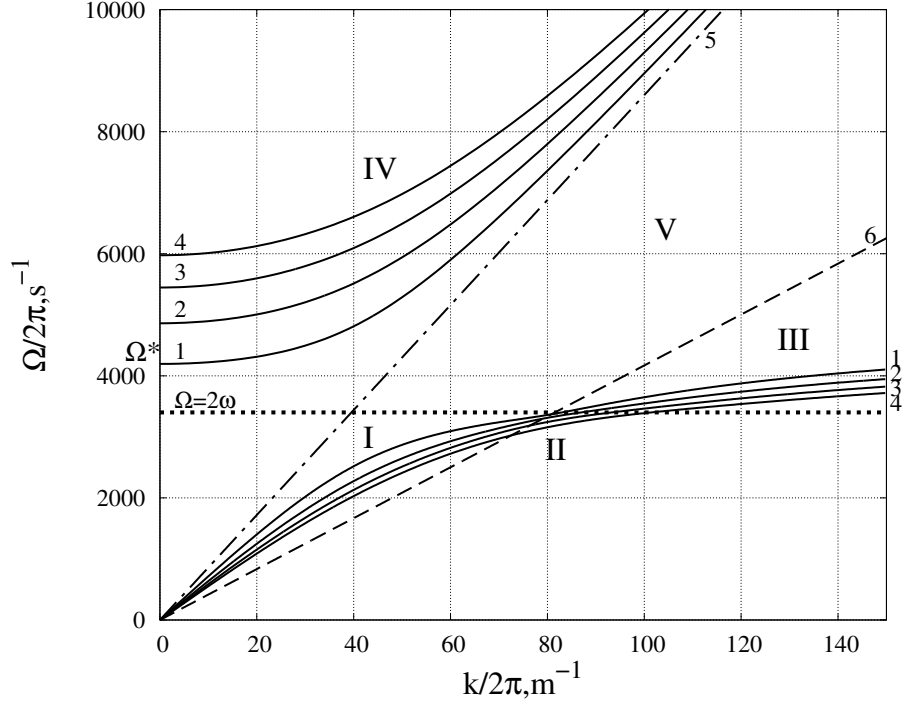}
\caption{Law of dispersion for first 4 radial modes of the upper and lower waves (lines 1-4). Line 5 - law of dispersion of the conventional sound waves $\Omega=ck$, line 6 - shows  the location where $A'=0$.}
\label{fig:dispersion}
\end{figure}

For the numerical calculations the parameters of the gas are given in table   \ref{tab:1}. 
Function $\Omega(k)$ is shown in fig. \ref{fig:dispersion}. For better  understanding of the dispersion equation it is convenient
to remind that eq. (\ref{form:eqosn})  can be interpreted as the  Shr\"{o}dinger equation for a particle in the potential of the form (\ref{shredinger}).
The lines where $A'$ and $B'$ equals to zero are shown in this figure.  Condition  $B'=0$ produces two lines. This equation is fulfilled when $\Omega=2\omega$ and on the line $\Omega=kc$, corresponding to the  law of dispersion of the conventional sound waves. The last line $\Omega={2\sqrt{\gamma-1}\over \gamma }ck$ corresponds to the condition $A'=0$. This line goes below the line $\Omega=kc$ for any real gas with adiabatic index $\gamma < 2$.  The working gas $UF^6$ has  $\gamma =1.067$.

\begin{table}
\caption{\label{tab:1}Basic parameters of GC Iguasu}
\begin{center}
\begin{tabular}{cc}

Parameter & Value  \\

$M$ & 352 g/mol\\
$a$ & 0.065 m \\
$\omega$ & $2\pi\times 1700 s^{-1}$ \\
$T_0$ & 300 K\\
$p_w$ & 80 mm Hg\\
$c_p$& 385 $J\cdot K/kg$ \\
$c$& 86 m/s\\
\end{tabular}
\end{center}
\end{table}

The waves split into two families in dependence of the phase velocity of the waves. Upper family  with $\Omega > kc$ and lower family  with $\Omega < kc$. 
Fig. \ref{fig:dispersion} shows only first 4 radial modes of the waves in dependence on the wave vector $k$ along the rotational axis. 
According to the figure the curves for upper family go to the line 5 at growth of $k$. Their phase velocities go to the sound velocity in the limit $k \rightarrow \infty$. The frequency of the waves increases with the number of the radial mode.  At $k\rightarrow 0$  the frequency of the first mode  $\Omega$ goes to $\Omega^*$ (fig. \ref{fig:dispersion}). In the case $\frac{M \omega^2 a^2}{2}\gg k T_0$ (conventional case) we have the following equation for $\Omega^*$
\begin{equation}
 \Omega^*=1.2335 \cdot 2 \omega.
\end{equation}

\begin{figure}
\centering
\includegraphics[width=0.99\linewidth]{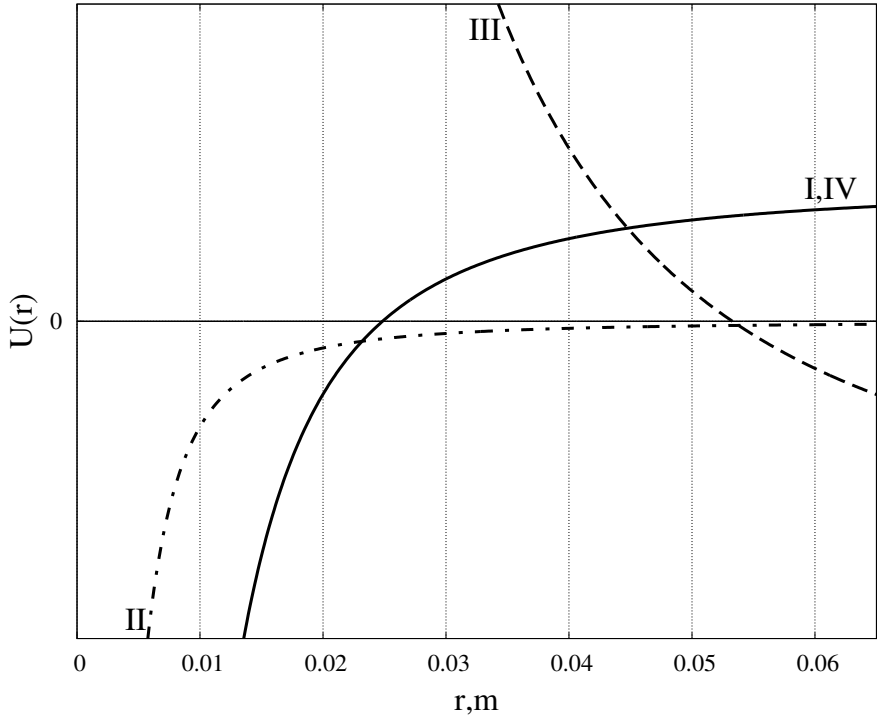}
\caption{Dependence of  U on r for  the waves of upper and lower families. Numbers near the  curves correspond to the zones where the potential is
realised. }
\label{fig:potential1}
\end{figure}

\begin{figure}
\centering
\includegraphics[width=0.99\linewidth]{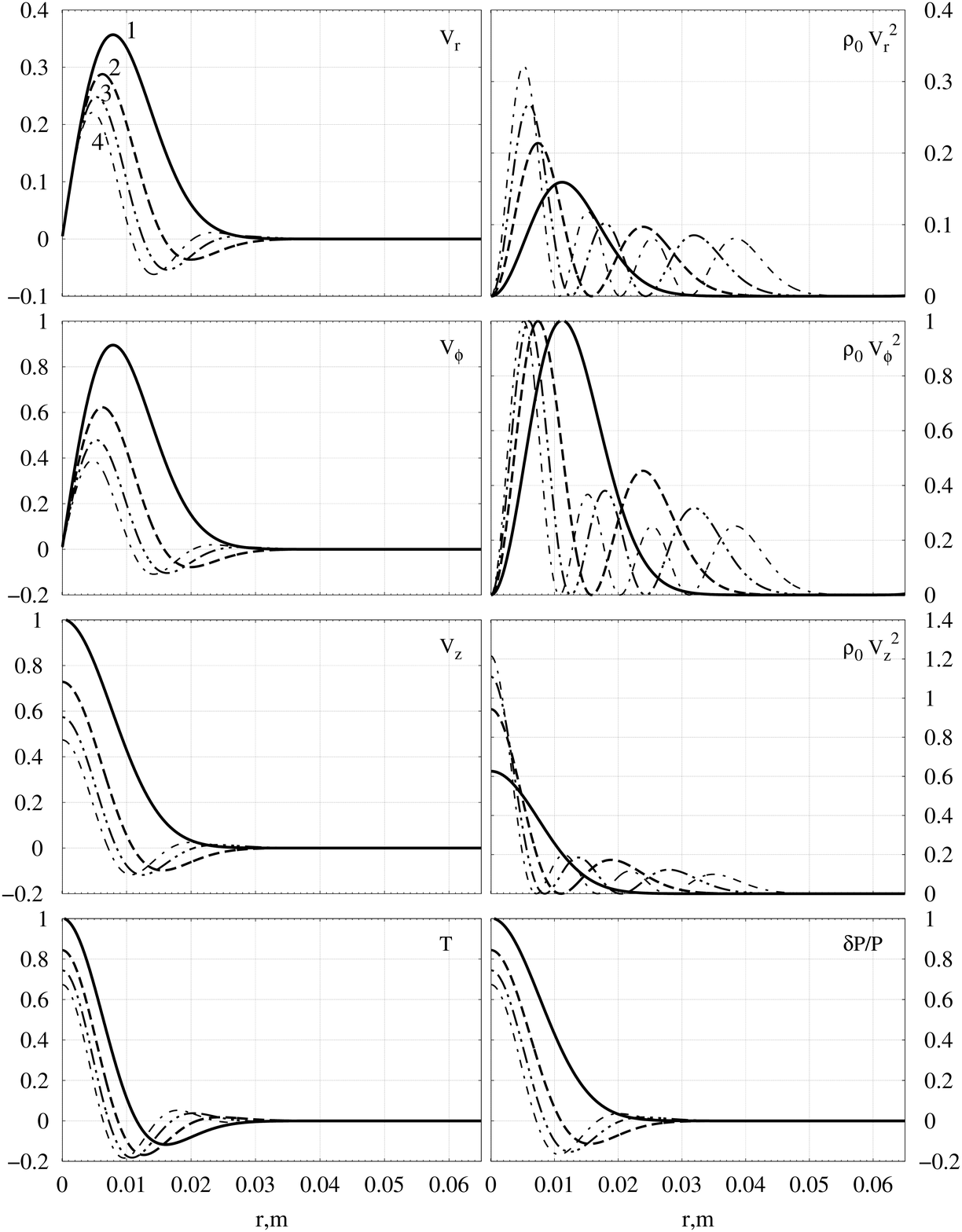}
\caption{Radial dependence of characteristics of the upper family of the waves at $k=15$. Numbers in the left upper panel correspond to the number of the radial mode. Components of the velocity  are normalized on the maximal value of one of the 
components. This component reaches in the maximum 1. Components of the energy density are normalized on the maximal value of one of the components of the energy density. This component of the kinetic energy density reaches 1 at the maximum. Thus, the actual  relationship between the components of the velocity and energy density is shown.}
\label{fig:whupk15}
\end{figure}

\begin{figure}
\centering
\includegraphics[width=0.99\linewidth]{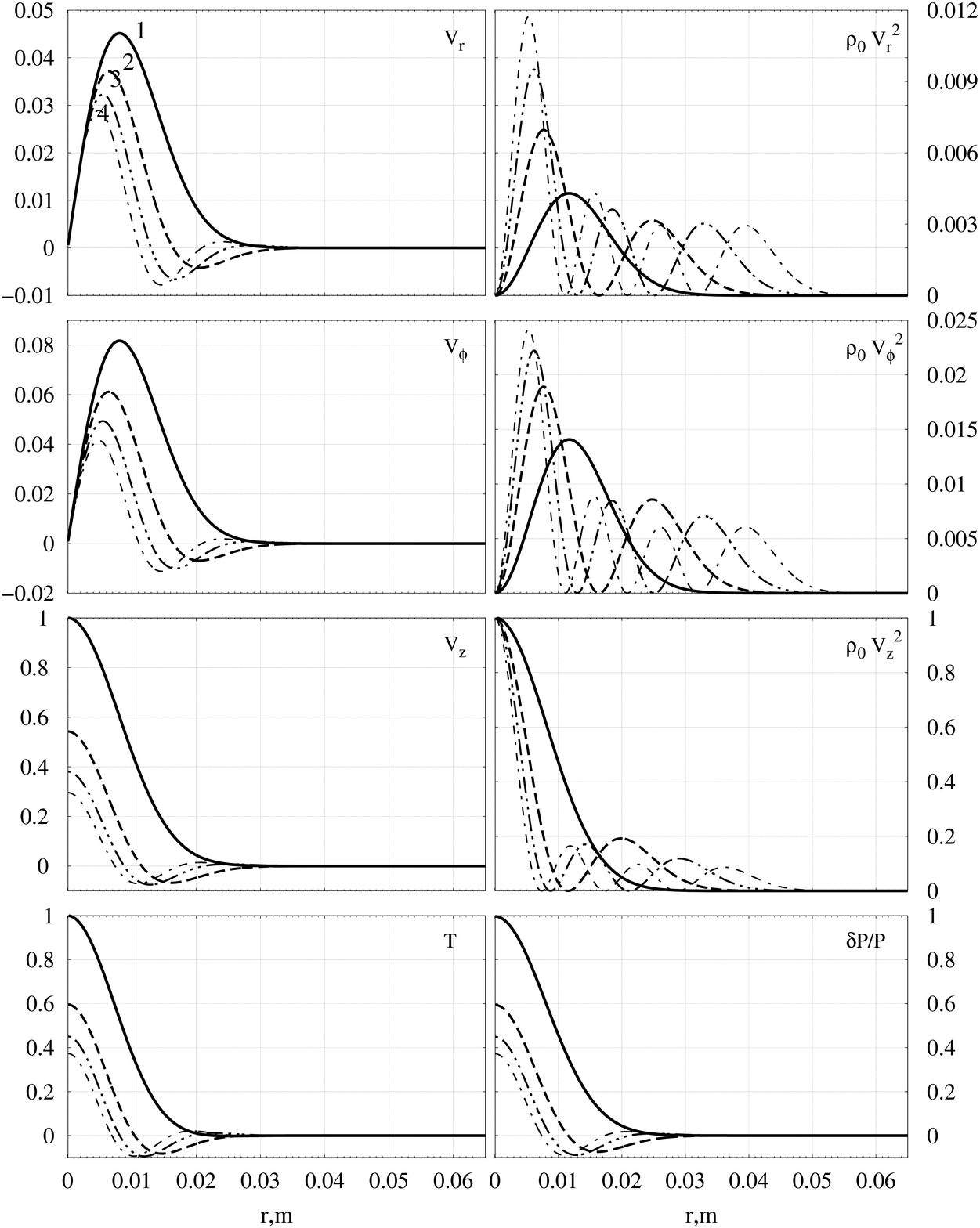}
\caption{Radial dependence of characteristics of the upper family of the waves at $k=60$. Numbers in the left upper panel correspond to the number of the radial mode. Components of the velocity  are normalized on the maximal values of one of the 
components. This component reaches in the maximum 1. Components of the energy density are normalized on the maximal value of one of the components of the energy density. This component of the kinetic energy density reaches 1 at the maximum. Thus, the actual relationship between the components of the velocity and energy density is shown.}
\label{fig:whupk60}
\end{figure}

For the upper family of the waves the potential $U(x)$ have a form presented in fig. \ref{fig:potential1}, curves I, IV. This kind of potential is typical for this family at any $k$ because $\omega > kc$,  $\Omega > 2\omega$ and $A' < 0$. We see that there is a  classical region in this potential where $U(x) \le 0$. The ``particle`` in this region can move  with energy equal to zero.  This region is located near the rotational axis. In the forbidden zone (close to the rotor wall)  the waves are exponentially damped.  This means that the perturbations of all variables will be concentrated close to the axis of rotation.  This conclusion is supported by the results of calculations of the radial distribution of the perturbations.  It follows from figs. \ref{fig:whupk15} and \ref{fig:whupk60} that the energy density of the perturbations is concentrated near the rotational axis. It is interesting that at small $k$ (the group velocity much less $c$) the energy difference between $r$, $z$ and azimuthal components of the velocity does not exceed factor 5 (fig. \ref{fig:whupk15}). Basically the kinetic energy is concentrated in the azimuthal component. However,  at large $k$ (the group velocity close $c$) the polarization of the waves becomes closer to the polarization of conventional longitudinal sound waves where the largest part of the energy is concentrated in the $z$ component of the velocity (fig. \ref{fig:whupk60}). The energy density in this component exceeds ones in other components on a few orders of magnitude. 

\begin{figure}
\centering
\includegraphics[width=0.99\linewidth]{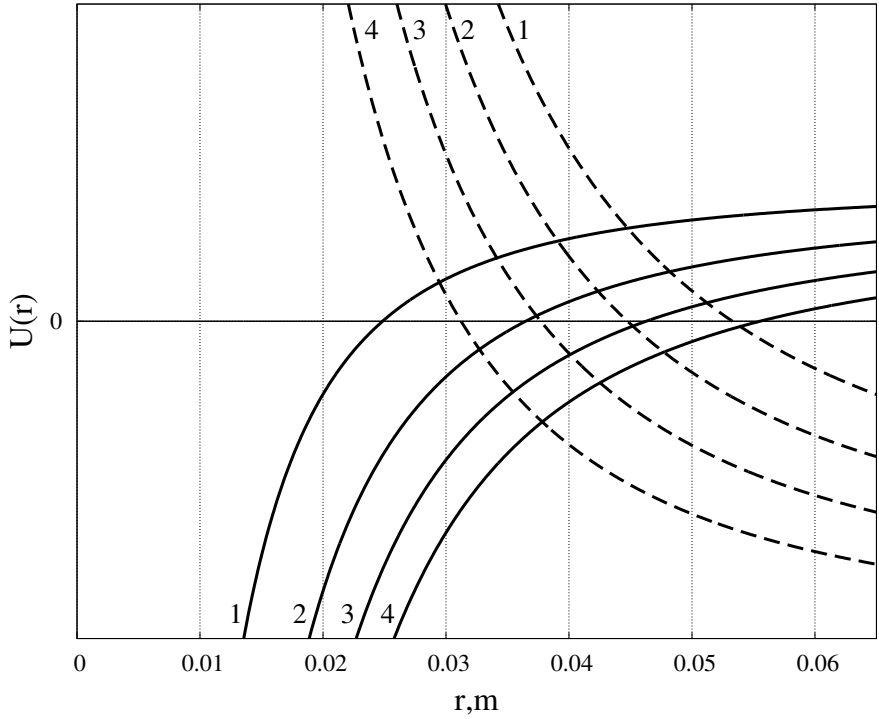}
\caption{Potential U(r) for different radial modes. Solid lines show potential for the zones I,IV, while  dashed lines for the zone III.}
\label{fig:potential2}
\end{figure}

\begin{figure}
\centering
\includegraphics[width=0.99\linewidth]{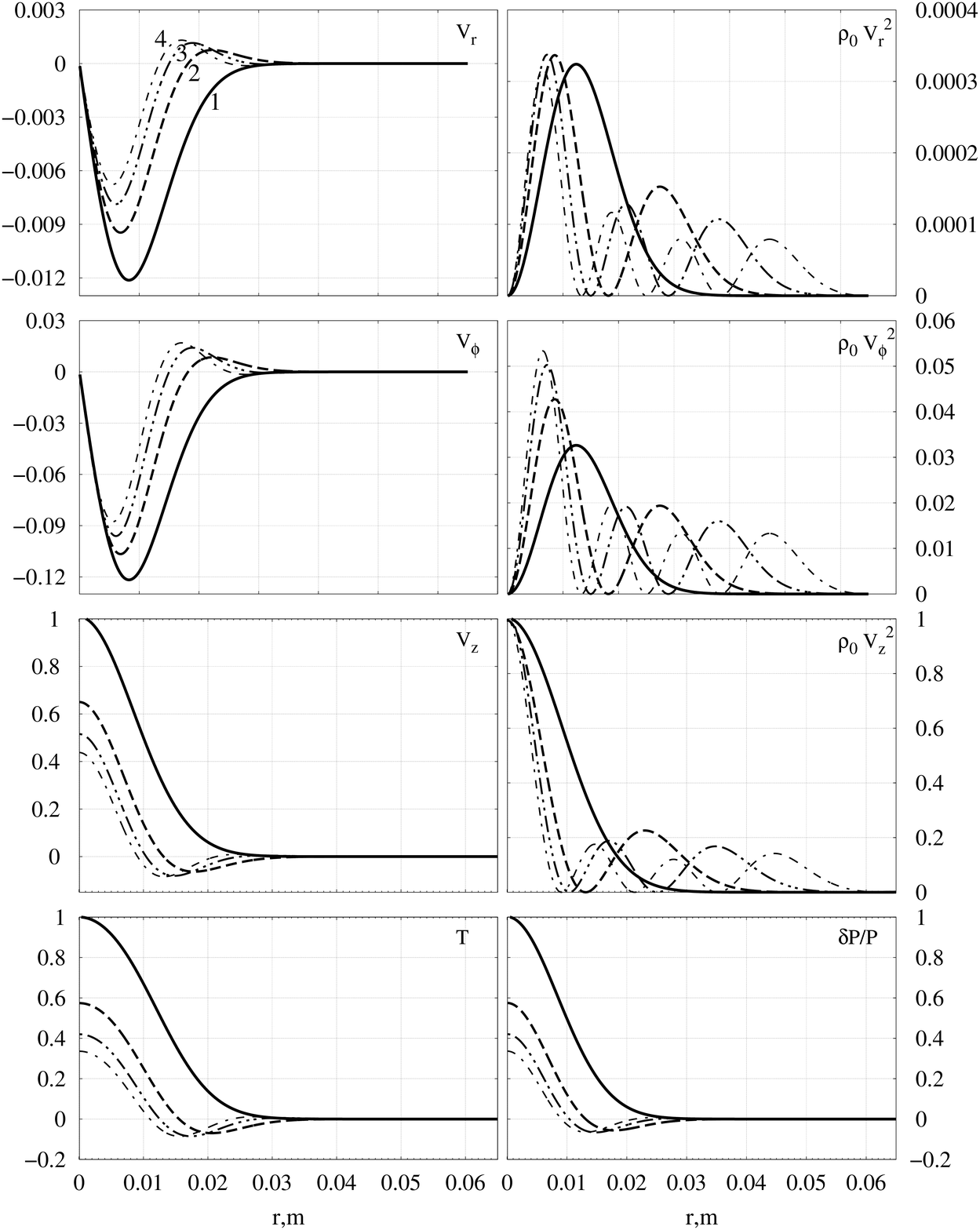}
\caption{Radial dependence of characteristics of the lower family of the waves at $k=15$}
\label{fig:whdownk15}
\end{figure}

The dispersion relation of the lower family of the curves is more complicated. Analogy of our problem with the Shr\"{o}dinger equation helps us to understand 
the physics of the dispersion curves. The potential changes form because $A'$ and $B'$ changes sign while we move along the curve with growth of $k$. 
 There are three zones. Number I marks the zone where $A< 0$ and $B>0$ in fig. \ref{fig:dispersion}. In this potential the perturbations concentrate at the rotational axis. Direct calculations presented in fig. \ref{fig:whdownk15}  confirms this conclusion. 

In the zone II $A'$ changes sign and the potential takes a form shown in fig. \ref{fig:potential1}. In this case  function $Y$ for the first mode of the radial oscillations will more or less uniform along radius. Nevertheless, the exponential factor in eqs. (\ref{eqvr}-\ref{eqvz}) will make the energy density concentrated at the axis.  

\begin{figure}
\centering
\includegraphics[width=0.99\linewidth]{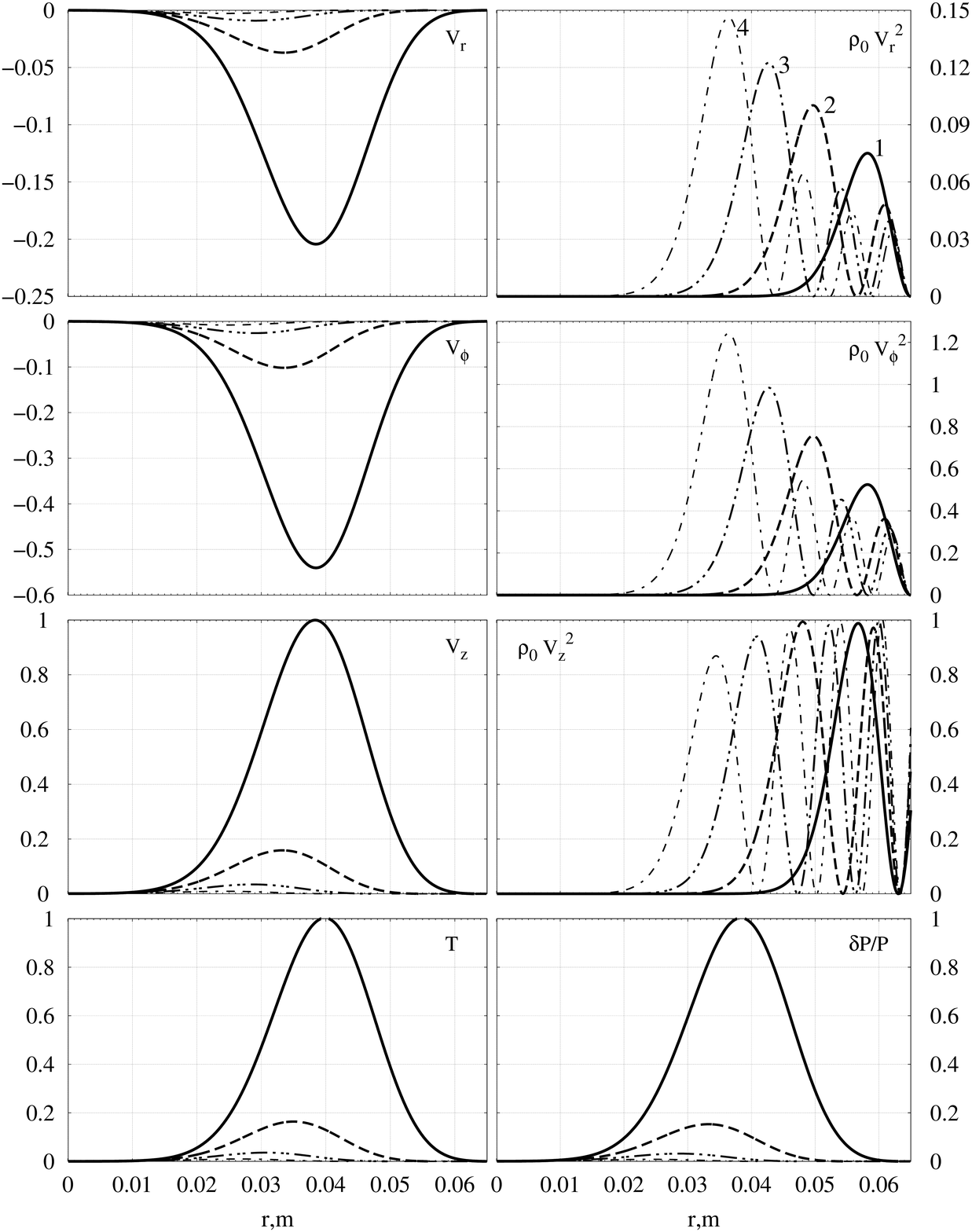}
\caption{Radial dependence of characteristics of the lower family of the waves at $k=140$}
\label{fig:whdownk140}
\end{figure}

In the last zone III both $A'> 0$ and $B'>0$. Potential $U(x)$ takes form shown in fig.   \ref{fig:potential1} by curve marked number III. Classical region for a ``particle`` is located near the wall of the rotor in this case although this region extends to the axis with growth of the number of the radial mode as it follows from fig.\ref{fig:potential2}.   In this case function $Y$ should differ from zero in a region near the wall of the rotor, and all the perturbations are concentrated basically near the wall of the rotor as it is shown in fig. \ref{fig:whdownk140}.   Comparison of the position of the maximum of the density of the kinetic energy of the waves with the potential presented in fig. \ref{fig:potential2} shows that  the increase of the mode number is due to the shift of the potential to the axis of rotation indeed. Correspondingly, the energy density of the waves is shifted to the axis as well.  

There is  one important general property of the waves of the upper and lower families. The energy density of these waves peaks in rather rarefied regions of the gas where density of the gas is small compared with the density at the wall of the rotor. Even for the case of lower family the maximum of the energy density of the waves is located at $\sim 1 $ cm from the rotor wall (see fig. \ref{fig:whdownk140}). These solutions will be strongly affected by two factors. Firstly, the molecular viscosity will dominate the dynamics of the gas in this region and modify the solution.  Secondly, the hydrodynamics can not be applied in the very rarefied regions close to the axis. Correct solution here can be 
obtained only in the kinetic approximation. Therefore in future these waves should be reconsidered taking into account at least molecular viscosity.

\begin{figure}
\centering
\subfigure[]{
\includegraphics[width=0.48\linewidth]{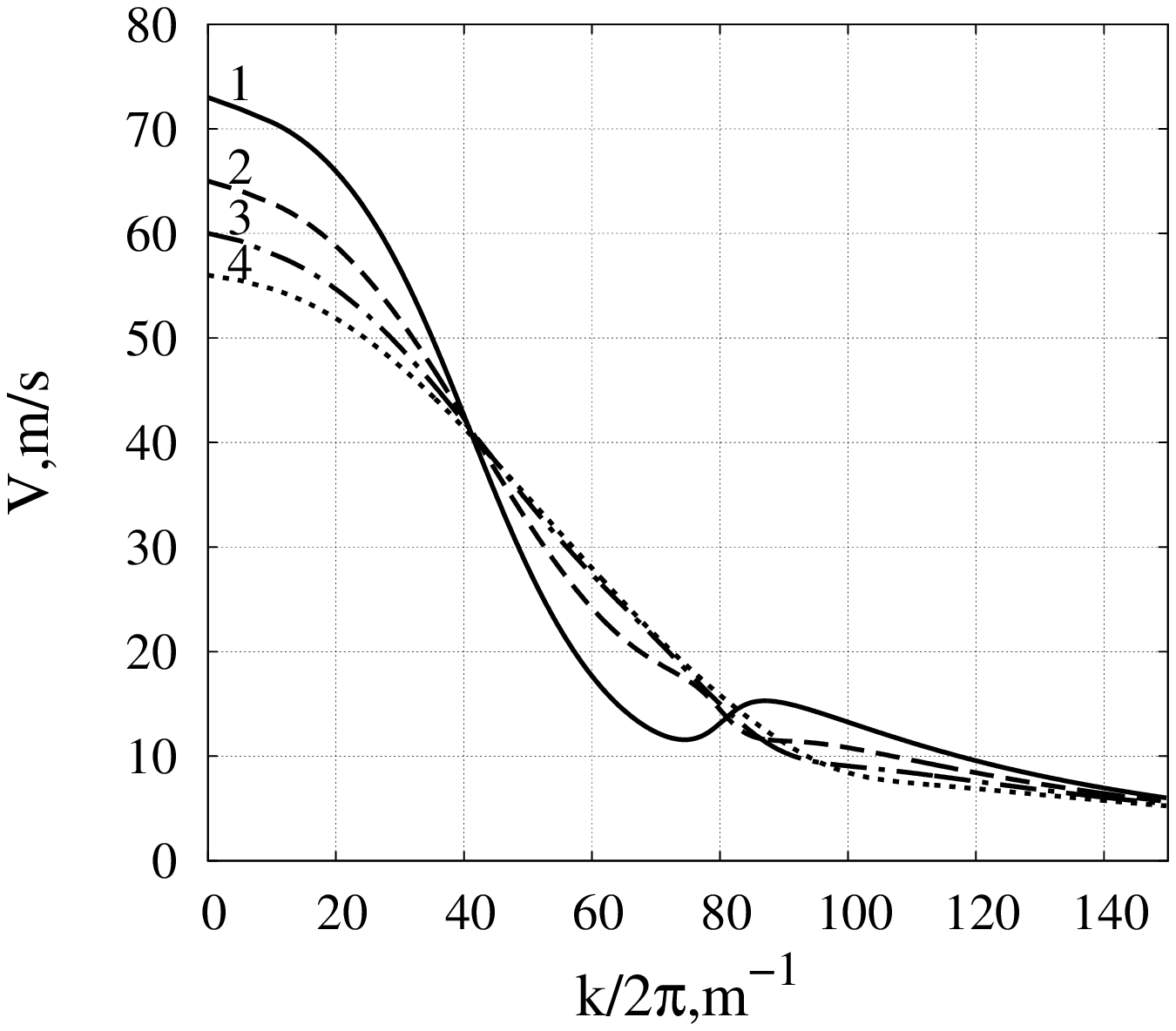}
\label{nizgr}
}
\subfigure[]{
\includegraphics[width=0.47\linewidth]{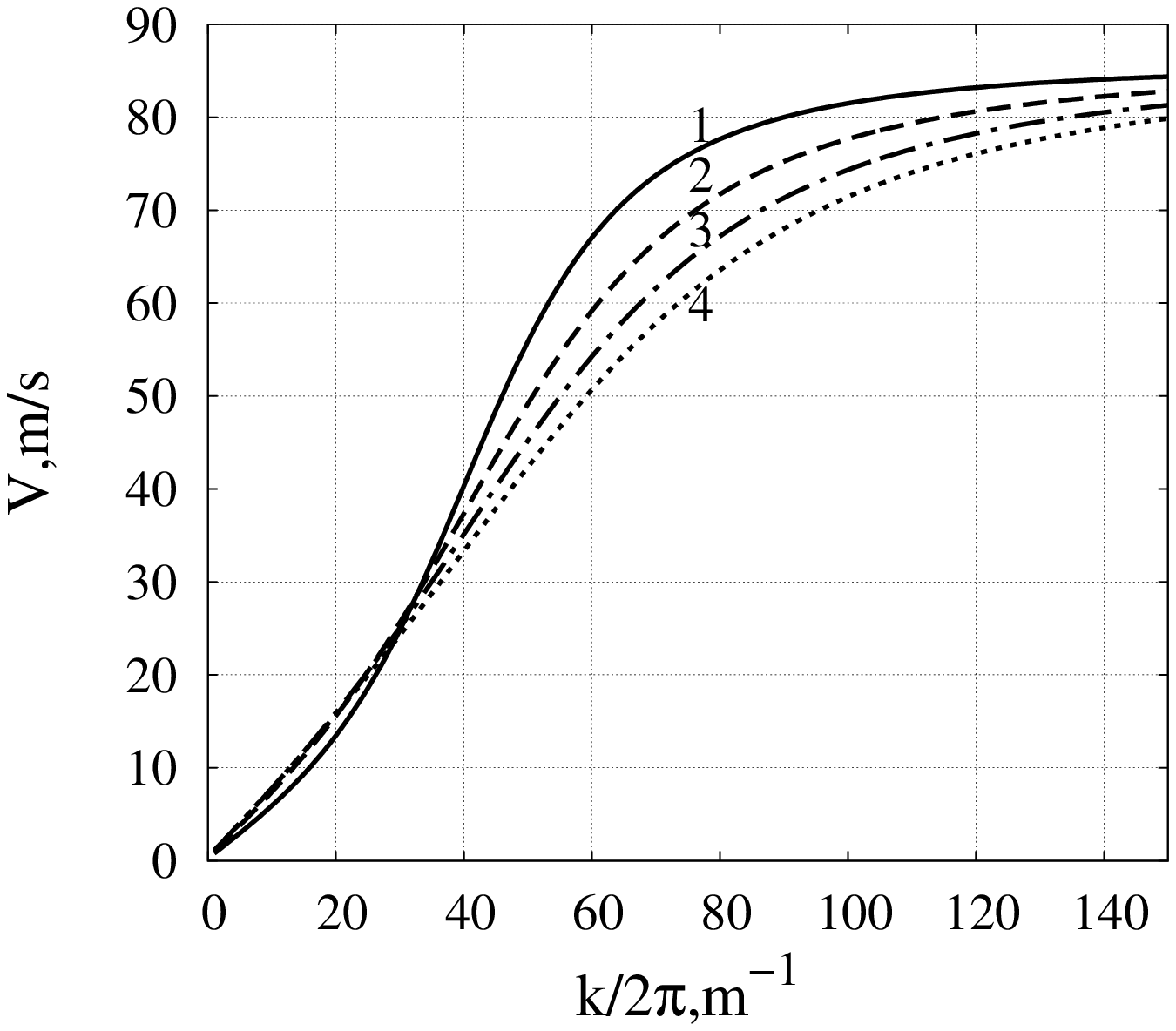}
\label{verhgr}
}
\caption{Group velocities of the waves:
\subref{nizgr} lower family;
\subref{verhgr} upper family;
}
\label{fig:velgr}
\end{figure}

\begin{figure}
\centering
\subfigure[]{
\includegraphics[width=0.48\linewidth]{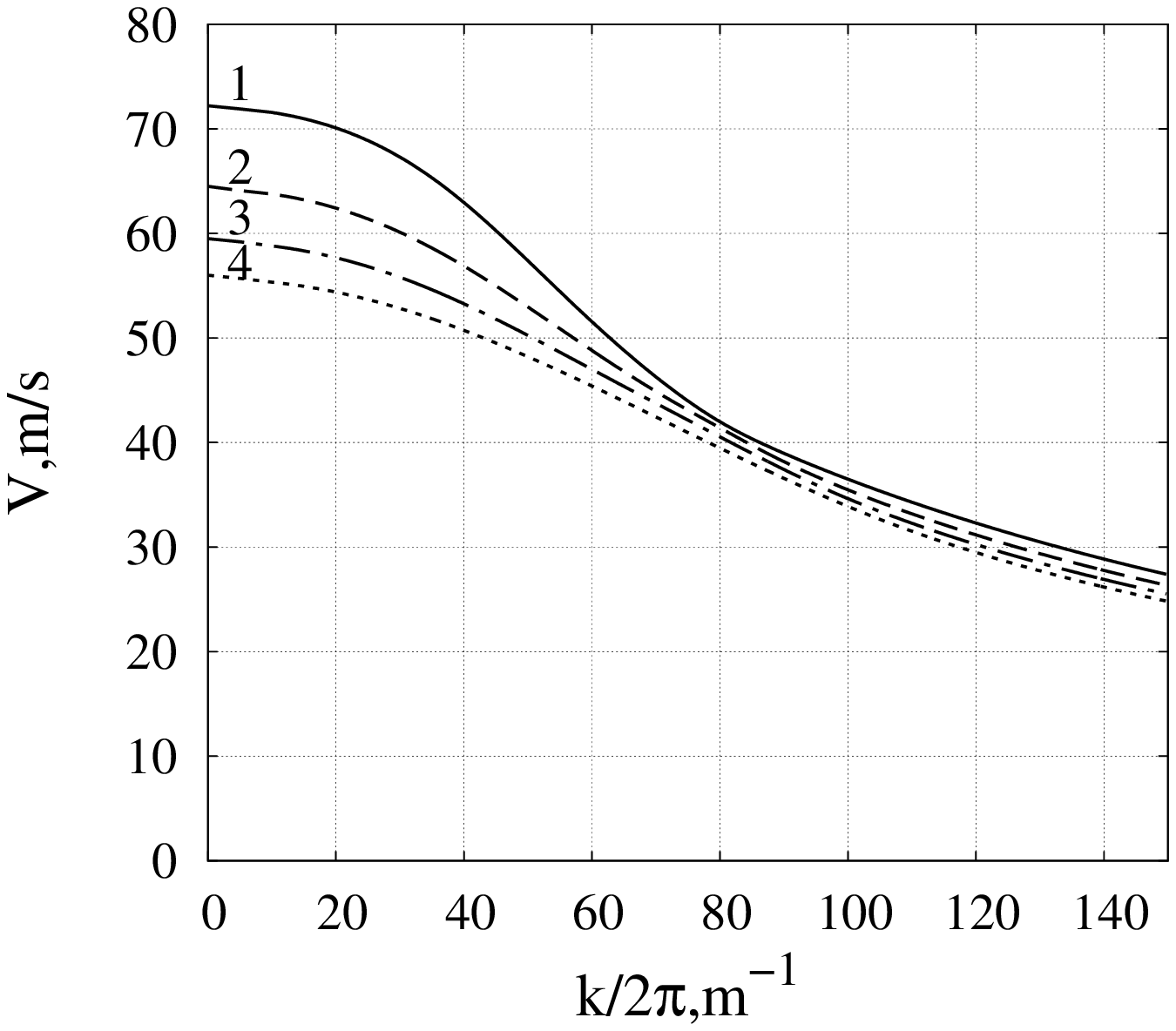}
\label{niz}
}
\subfigure[]{
\includegraphics[width=0.47\linewidth]{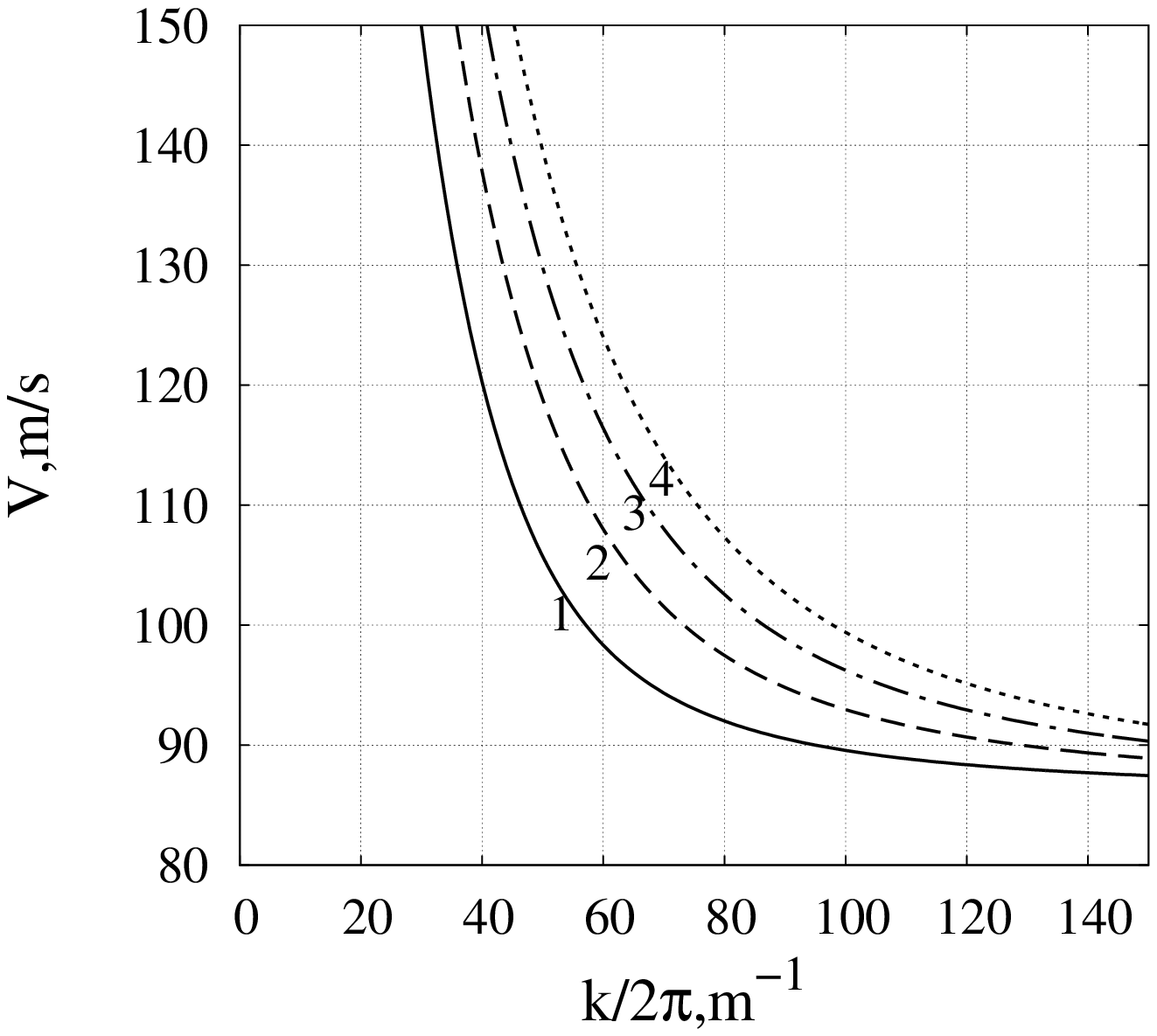}
\label{verh}
}
\caption{Phase velocity of the waves:
\subref{nizph} Lower family;
\subref{verhph} Upper family;
}
\label{fig:velph}
\end{figure}

Figs. \ref{fig:velgr}, \ref{fig:velph} show the phase and group velocities of the waves. They essentially depend on the wave length and strongly differs from the conventional sound velocity. Only phase and group velocities of the upper family of the waves converges to the sound velocity at $k\rightarrow \infty$.
The most interesting feature is the non monotonic behaviour of the group velocity of the first mode of the lower family of the waves.  This behaviour is connected with the transition through the lines $A'=0$ and $\Omega=2\omega$. No waves can exist in the zone V. Potential $U(r)$ is above 0 here. Therefore any dispersion curve should go below the point where the lines $A'=0$ and $\Omega=2\omega$ cross each other. The potential changes here quickly from type I to type III.  This deforms  the dispersion curves and this deformation results into non monotonic behaviour of the group velocity of the first mode.

\subsection{Case $A=0$}

\begin{figure}
\centering
\subfigure[]{
\includegraphics[width=0.48\linewidth]{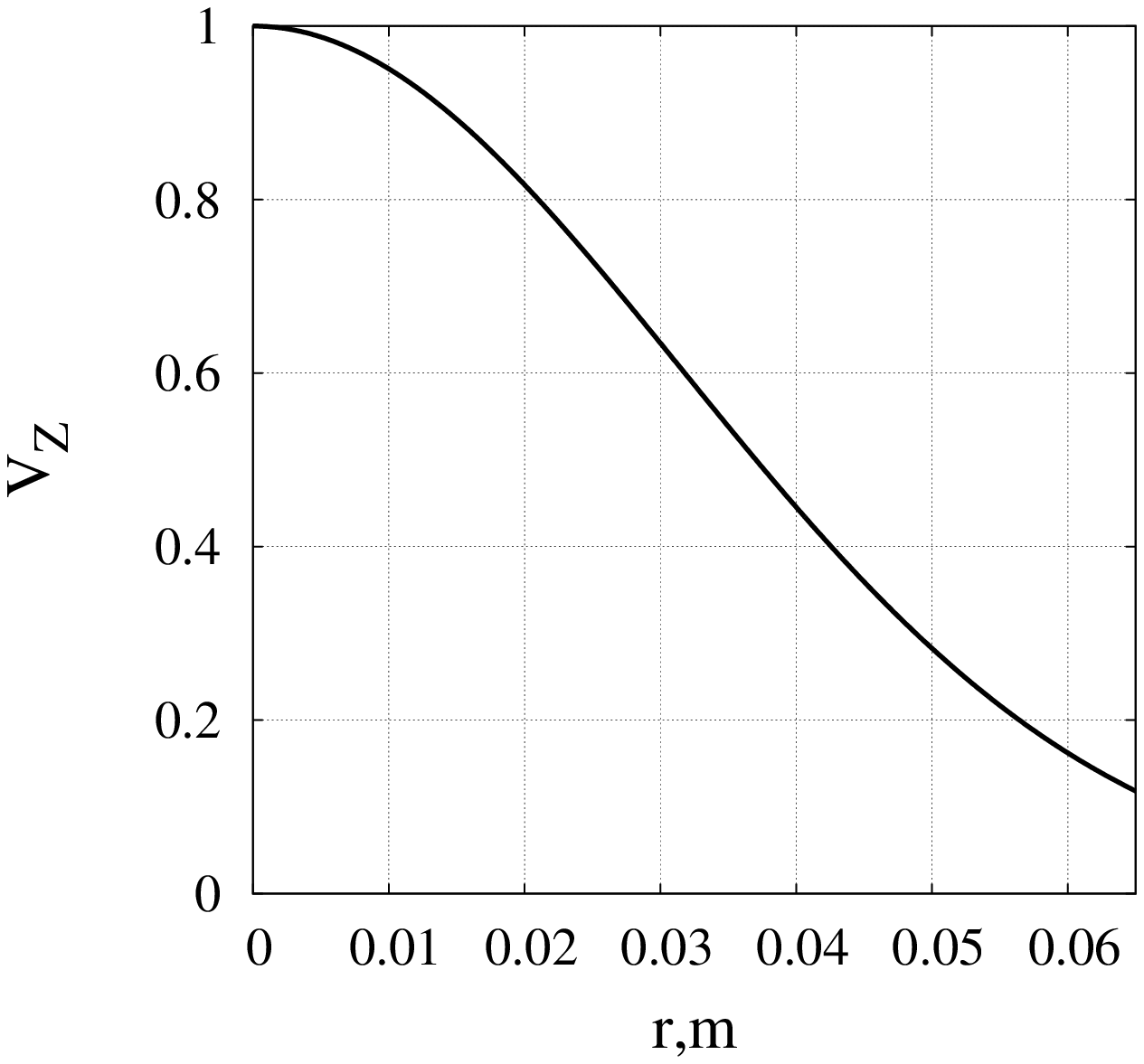}
\label{nizph}
}
\subfigure[]{
\includegraphics[width=0.47\linewidth]{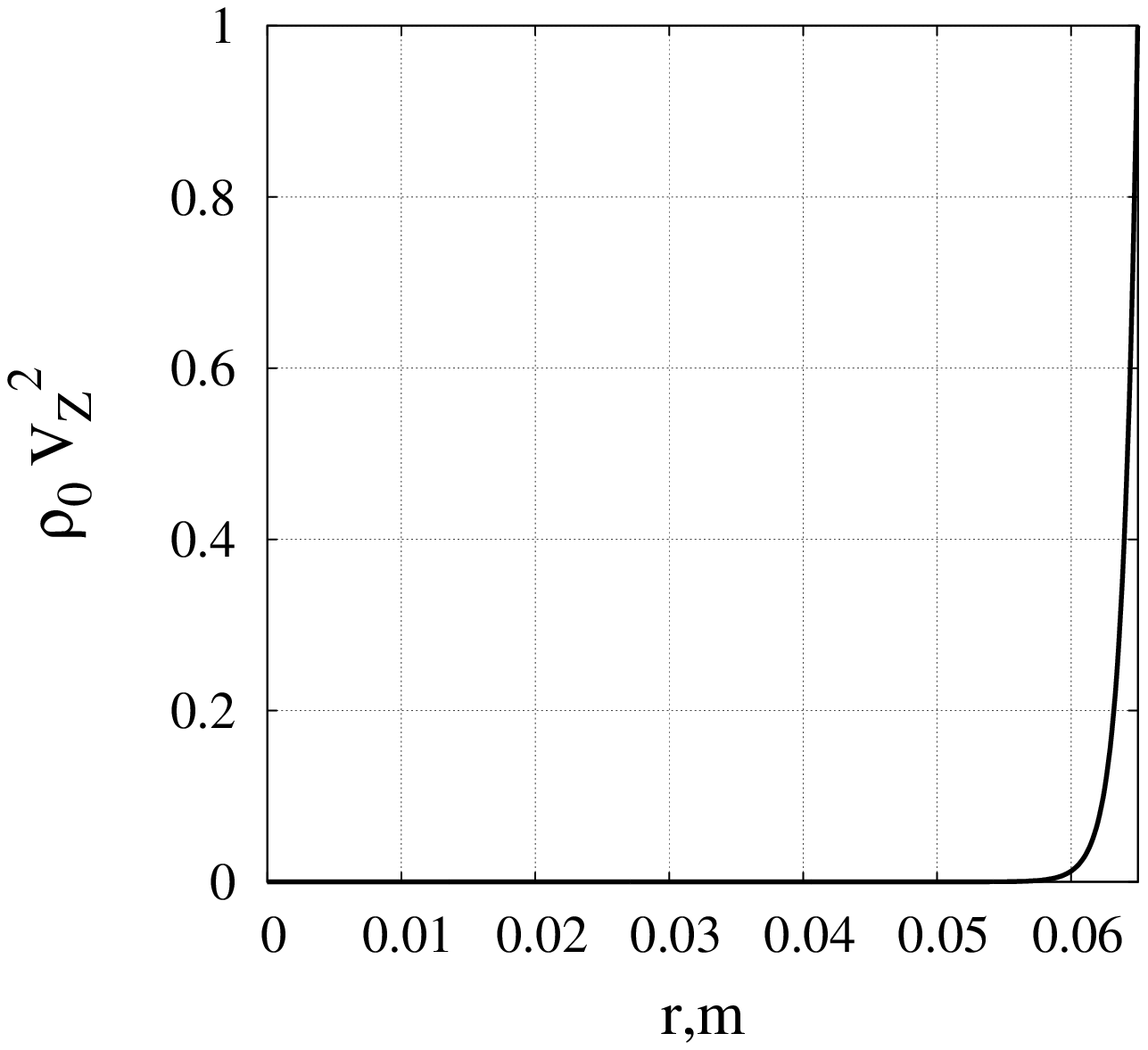}
\label{verhph}
}
\caption{Radial distribution in sound family of the waves of:
\subref{nizph} velocity;
\subref{verhph} density of the kinetic energy;
}
\label{sound}
\end{figure}

The case when $A=0$ needs  separate consideration  because the solution corresponding to this condition has been lost  when eq.  (\ref{form:pres}) has been obtained. This family of the waves propagate with the velocity of conventional sound waves. The dispersion relation has a form $\Omega=kc$. This family of the waves we will call as sound family. This is strictly longitudinal wave.  $V_r=0$ and $V_{\varphi}=0$ in this case. Equation for the pressure perturbation can be obtained from right hand part of eq. (\ref{form:eq1}). It has a form
\begin{equation}
\frac{\omega^2 r p}{c^2}-\frac{\partial p}{\partial r} =0.
\end{equation}
The solution of this equation is trivial
\begin{equation}
 p=\bar p_w \exp{\frac{\omega^2(r^2-a^2)}{2c^2}},
\end{equation}
where $\bar p_w$ is the perturbation of the pressure at at the wall of the rotor. 
$V_z$ is connected with the perturbation of pressure by eq. (\ref{eqvz}). It follows from this equation that
\begin{equation}
V_z={c\over \gamma}{\bar p_w\over p_w} \exp{\frac{(1-\gamma)\omega^2(r^2-a^2)}{2c^2}}.
\label{soundvz}
\end{equation}
Keeping in mind that $\gamma =1.067$ for $UF^6$ it is clear that the perturbation of the velocity is distributed along radius much more uniformly compared with the distribution of pressure or  pressure perturbation in this case. 
Distribution of the velocity and kinetic energy  in the wave is presented in fig. \ref{sound}.
It follows from this figure that the velocity decreases with radius only on a factor of 5  while the kinetic energy of the wave changes on 12 -14 orders of magnitude and is strongly concentrated in the narrow region near the 
wall of the rotor. This property essentially distinguish this family of the waves from the waves discussed in the previous subsection.

\section{Discussion}

The basic characteristics of the linear waves in the gas in strong centrifugal field were considered in hydrodynamical approximation neglecting viscosity and heat conductivity. Conventional  acoustic waves split into three families in the strong centrifugal field.  Spatial distribution, polarisation and  velocity of the waves  essentially differ from the conventional acoustic waves. The families of the waves called by us as  upper and lower families concentrate energy close to the axis of rotation where the gas is so strongly rarefied that  the hydrodynamic approximation is not valid here. It is evident that these types of the waves should be considered in more accurate approximation. Now it is even not clear what kind of approximation should be used. This is a challenge for theoreticians to develop mathematical methods for 
accurate description of the waves in the conditions of so strong centrifugal fields.    Nevertheless, even now it is clear that these waves will be strongly dumped due to dissipative processes because the viscosity and thermal conductivity dominate all other processes in so rarefied gas. 

  The sound family of the waves is the most  interesting for the physics of gas flows in  GC.  The energy density of these waves concentrates near the wall of the rotor. On this reason dumping decrement for this family  is expected to be the lowest. The waves propagate for a longer  distance in compare  with  the waves of other families. 
Violation of the hydrodynamic approximation near the axis has no importance for these waves because all the energy is concentrated in the region where this approximation is valid. However, the dumping decrement of this family of the waves can not be estimated directly from known formulas for damping of acoustic waves \citep{landau} on a few reasons. Firstly, the distribution of the energy density of the waves and velocity strongly modify the decrement. Remind, that the energy density is concentrated near the walls, while the velocity of the gas is localized closer to the axis of rotation. Secondly, the friction and heat exchange with the wall of the rotor modify the decrement as well. Thus, the solution of the problem of definition of the dumping of the linear waves will demand development new mathematical methods which will take into account all the most important features of the dumping process.

\section{Acknowledgement}
The authors are grateful to Drs. V.Borman, V. Borisevich  and V. Tronin for useful discussions.  The work has been performed under support of  Ministry of education and science of Russia, grant no. 3.726.2014/K

\end{document}